\documentclass[pre,amssymb,twocolumn,raggedbottom,nofootinbib]{revtex4}
     \usepackage{graphicx}
\begin{document}
\renewcommand{\thefootnote}{\fnsymbol{footnote}}
\title{Local molecular field theory for effective attractions between like charged objects in systems with strong Coulomb interactions}
\author{Yng-Gwei Chen$^{1,3,*}$ and John D. Weeks$^{2,3}$}
\affiliation{%
$^{1}$Department of Physics, $^{2}$Department of Chemistry and Biochemistry,
and $^{3}$Institute for Physical Science and Technology,
University of Maryland, College Park, Maryland 20742}
\maketitle

{\bf Strong short ranged positional correlations involving counterions can
induce a net attractive force between negatively charged strands of DNA, and
lead to the formation of ion pairs in dilute ionic
solutions. But the long range of the Coulomb interactions impedes
the development of a simple local picture.
We address this general
problem by mapping the properties of a nonuniform system with
Coulomb interactions onto those of a simpler system with short
ranged intermolecular interactions in an effective external field
that accounts for the averaged effects of appropriately chosen long
ranged and slowly varying components of the Coulomb interactions. The
remaining short ranged components combine with the other molecular core
interactions and strongly affect pair correlations in dense or
strongly coupled systems. We show that
pair correlation functions in
the effective short ranged system closely resemble those in the uniform
primitive model of ionic solutions, and illustrate the formation of ion
pairs and clusters at low densities. The theory accurately describes
detailed features of the effective attraction between two equally
charged walls at strong coupling and intermediate
separations of the walls. New analytical results for
the minimal coupling strength needed to get any attraction and for the
separation where the attractive force is a maximum are presented.}

Strong Coulomb interactions in crowded nonuniform environments have
important experimental consequences in a wide variety of biophysical
applications ranging from DNA packaging in viruses to transport in ion
channels \cite{gelbart00,levin02,grosberg02,boroudjerdi05}.
They present major challenges to theory and computer simulations not only
because of their characteristic long range but also because they can be
very strong at short distances.
We present here a new local molecular field (LMF) theory \cite{weeks02}
that averages over particular long ranged and slowly varying components
of the Coulomb interactions \cite{chen04} while still maintaining an accurate
description of the short ranged components.
It provides a general and physically suggestive theory for strongly
coupled Coulomb systems and reduces
exactly to the classical Poisson-Boltzmann (PB) approximation
for dilute weakly coupled systems.

We consider a general starting point where a molecule of species $i$,
described by a rigid body frame with center at ${\bf r}_{i}$, interacts with
an external field $\phi _{fi}({\bf r}_{i})$ that comes from fixed charged solutes, or
walls, or particular fixed molecules of a mobile species, as illustrated below.
The subscript $f$ indicates the source of the field, which we treat
as a special fixed species $f$. 
The interaction between a pair of molecules of species $i$ and $j$
is assumed to have the general form $w_{ij}({\bf r}_{ij})=w_{s,ij}({\bf r}%
_{ij})+w_{q,ij}({\bf r}_{ij}),$ where ${\bf r}_{ij}\equiv {\bf r}_{j}-{\bf r}%
_{i}.$ The $w_{s,ij}({\bf r}_{ij})$ denote general (repulsive core and
other) short ranged intermolecular interactions.
There are angular coordinates expressing orientations of the body frames
that we do not denote explicitly. The $w_{q,ij}({\bf r}_{ij})$
arise from Coulomb interactions between rigid charge distributions $q_{i}(%
{\bf r-}{\bf r}_{i})$ in the body frame of each molecule, so that 
\begin{eqnarray}
w_{q,ij}({\bf r}_{ij}) &=&\int d{\bf r}\int d{\bf r}^{\prime }\frac{q_{i}(%
{\bf r-r}_{i})q_{j}({\bf r}^{\prime }{\bf -r}_{j})}{\epsilon |{\bf r}-{\bf r}%
^{\prime }|}  \label{wqdefr} \\
&=&\frac{1}{(2\pi )^{3}}\int d{\bf k}\hat{q}_{i}(-{\bf k})\hat{q}_{j}({\bf k}%
)e^{-ik\cdot {\bf r}_{ij}}\frac{4\pi }{\epsilon k^{2}},  \label{wqdefk}
\end{eqnarray}
where the caret denotes a Fourier transform and we assume there is a uniform
dielectric constant $\epsilon $ everywhere.

\footnotetext{%
Abbreviations: LMF, local molecular field; PB Poisson-Boltzmann; SAPM
size asymmetric primitive model; WL Weis and Levesque; MC Monte Carlo;
SCA strong coupling approximation; mPB mimic Poisson-Boltzmann
\\
$^*$ Present address: Laboratory of Chemical Physics and National Institute of Diabetes
and Digestive and Kidney Diseases, Bldg. 5, National Institutes of Health, Bethesda, MD 20892 }

To generate uniformly slowly varying components $u_{1,ij}$ of the full
$w_{q,ij}\equiv w_{q0,ij}+u_{1,ij}$ that are well suited for LMF averaging,
we limit the magnitude of wave vectors making significant contributions to
the integration in Eq.~\ref{wqdefk}. To that end we introduce a Gaussian
function parameterized by an important length scale $\sigma $ that provides
a smooth cutoff in $k$-space, and write 
\begin{equation}
\frac{4\pi }{k^{2}}=\frac{4\pi }{k^{2}}e^{-\frac{1}{4}(k\sigma )^{2}}+\frac{4%
\pi }{k^{2}}(1-e^{-\frac{1}{4}(k\sigma )^{2}}).  \label{coulombsepk}
\end{equation}
The first term on the right has all the characteristic long ranged Coulomb
divergences as $k\rightarrow 0$, but decays very rapidly to zero for
$k\sigma \gtrsim 2$. The desired slowly varying
components arise when only this term
is used in Eq.~\ref{wqdefk} with an appropriate choice of
$\sigma $. For localized charge distributions $\hat{q}_{i}({\bf k})$
we expand in a Taylor series about ${\bf k}=0$ and take the lowest order
multipole moment \cite{chen04a}. This
simplified expression defines the $u_{1,ij}$ we consider and thus prescribes
a $\sigma $-dependent separation of the full intermolecular potentials
$w_{ij}({\bf r}_{ij})=w_{s,ij}({\bf r}_{ij})+w_{q,ij}({\bf r}_{ij})=
w_{s,ij}({\bf r}_{ij})+w_{q0,ij}({\bf r}_{ij})+u_{1,ij}({\bf r}_{ij})\equiv u_{0,ij}({\bf r}_{ij})+u_{1,ij}({\bf r}_{ij})$ into
short and long ranged parts.

\begin{figure}[tbp]
\includegraphics[%
 width=1.0\columnwidth]{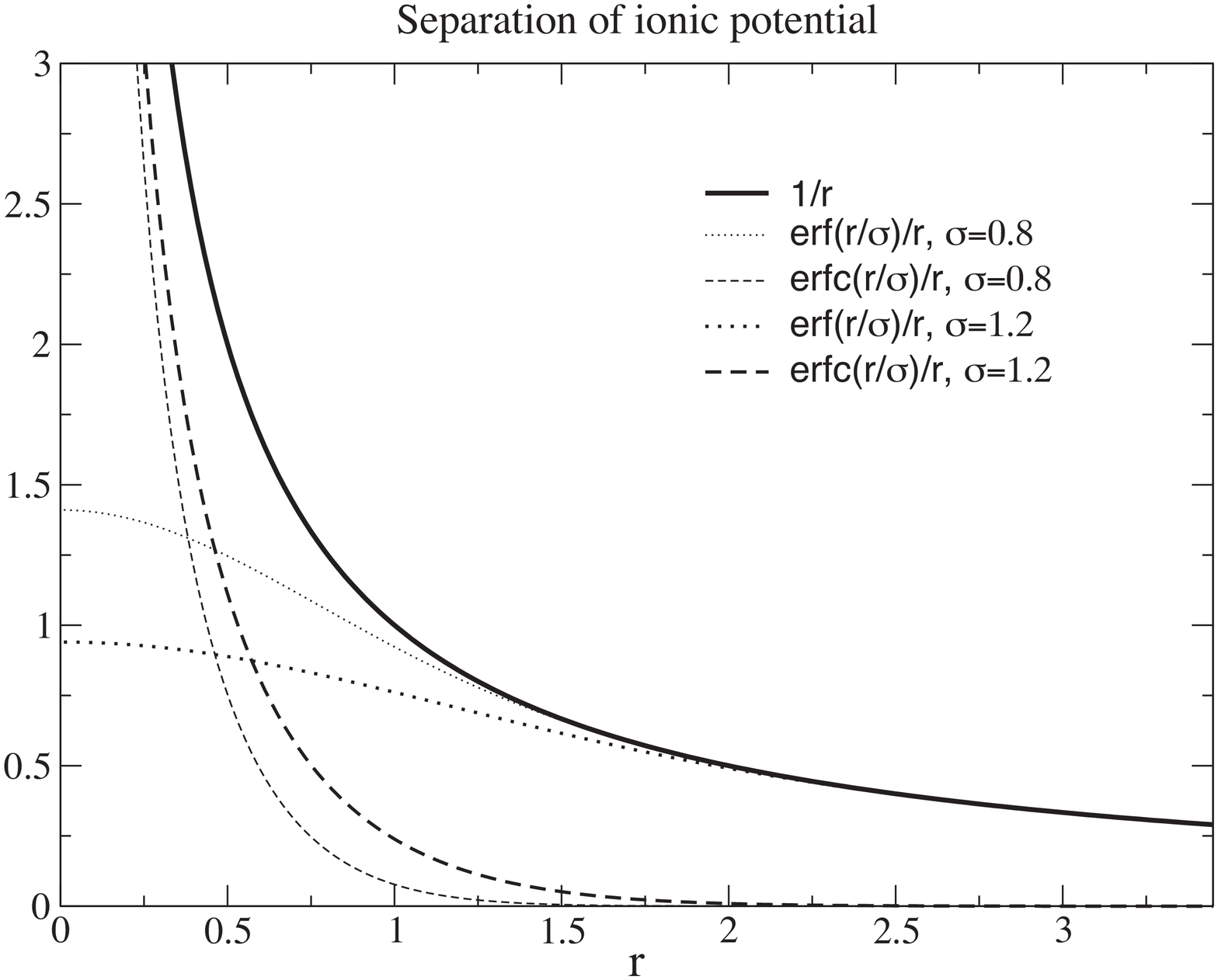}
\caption{Separation of the $1/r$ potential into a slowly varying piece
$\text{erf}(r/\sigma )/r$  and the short ranged remainder $\text{erfc}(r/\sigma )/r$. 
Two relevant $\sigma $ values are shown; a bigger $\sigma $ generates a more slowly
varying long ranged component.}
\label{potentialseparation}
\end{figure}

In $r$-space Eq.~\ref{coulombsepk} becomes $1/r=\text{erf}(r/\sigma )/r+%
\text{erfc}(r/\sigma )/r.$ Here erf and erfc are the usual error and complementary
error functions. The erf$(r/\sigma )/r$ term is the electrostatic potential
from a normalized Gaussian charge distribution with width $\sigma$. As
shown in Fig.~\ref{potentialseparation}, this remains
smooth and slowly varying on the scale of
$\sigma $, while decaying as $1/r$ at large $r$. 
This use of a Gaussian charge distribution is related to the Ewald
sum method, which considers periodic images of ion configurations
with embedded screening and compensating Gaussian charge distributions.
However, our focus is on the separation of the potential itself and not the
effects of periodic boundary conditions and our choice of ${\sigma }$ is
usually much smaller than that used in Ewald sum methods, which typically is
proportional to the width of the simulation cell \cite{frenkel02}.

The short ranged components $u_{0,ij}({\bf r}_{ij})$ define the
intermolecular interactions in the special short ranged ``mimic system".
They are comprised of the
short ranged parts of the Coulomb interactions $w_{q0,ij}\equiv
w_{q,ij}-u_{1,ij}$ and the other short ranged core interactions $w_{s,ij}$: 
\begin{equation}
u_{0,ij}({\bf r}_{ij})\equiv w_{s,ij}({\bf r}_{ij})+w_{q0,ij}({\bf r}_{ij}).
\label{u0mimic}
\end{equation}
As suggested by Fig.~\ref{potentialseparation}, $\sigma $ sets
the range of $w_{q0,ij}$ and thus
determines an effective Coulomb core size \cite{chen04}. The external
potential from fixed charged solutes or walls $\phi _{fi}({\bf r}_{i})\equiv 
$ $\phi _{0,fi}({\bf r}_{i})+\phi _{1,fi}({\bf r}_{i})$ can similarly be
separated into short and long ranged parts, as illustrated below.

It is straightforward to arrive at explicit results for the $u_{1,ij}({\bf r)%
}$ and $\phi _{1,fi}({\bf r}_{i})$. Here we consider
localized charge distributions with a net charge $\bar{q}_{i}\equiv \int d%
{\bf r}q_{i}({\bf r})$ or a net dipole moment ${\bf p}_{i}\equiv \int d{\bf %
rr}q_{i}({\bf r})$. If both molecules carry a net charge we find 
$u_{1,ij}(r)=\bar{q}_{i}\bar{q}_{j}\text{erf}(r/\sigma )/\epsilon r$.
The associated Coulomb core component is $w_{q0,ij}=\bar{q}_{i}\bar{q}_{j}
\text{erfc}(r/\sigma )/\epsilon r$. Thus the results of Fig.~\ref{potentialseparation},
scaled by $\bar{q}_{i}\bar{q}_{j}/\epsilon $, give examples of possible $u_{1,ij}$
and $w_{q0,ij}$ for ionic solution models \cite{chen04}. For a monopole and a
dipole we find $u_{1,ij}({\bf r})=\bar{q}_{i}({\bf p}_{j}\cdot \nabla )[\text{erf}
(r/\sigma )/\epsilon r],$ and for dipoles $u_{1,ij}({\bf r})=-({\bf p}_{i}%
\cdot \nabla )({\bf p}_{j}\cdot \nabla )[$erf$(r/\sigma )/\epsilon r]$. The
latter will lead to dipolar mimic systems with short ranged angular
dependent interactions.\\
\\
{\bf Local molecular field approximation}

LMF theory introduces renormalized effective fields $\phi _{R,fi}({\bf r}%
_{i})$ that induce nonuniform singlet densities in the mimic system
(denoted by the subscript $R$) that are supposed to equal those induced
by the $\phi _{fi}({\bf r}_{i})$ in full system of interest: 
\begin{equation}
\rho _{R,fi}({\bf r}_{i})=\rho _{fi}({\bf r}_{i}).  \label{rho0=rho}
\end{equation}
This defines a general mapping relating structure in the mimic and full
systems. Thermodynamic properties can be determined by integration of these
structural relations.

We represent the effective field $\phi _{R,fi}({\bf r}_{i})\equiv \phi
_{0,fi}({\bf r}_{i})+\phi _{R1,fi}({\bf r}_{i})$ as the sum of the known
short ranged part $\phi _{0,fi}({\bf r}_{i})$ of the external field in the
full system and a renormalized ``perturbation component'' $\phi _{R1,fi}(%
{\bf r}_{i})$ that accounts for the averaged effects of the slowly varying
interactions $u_{1,ij}$. As discussed in detail in \cite{chen04,weeks02}, by
considering the balance of forces in the full and mimic
systems when Eq.~\ref{rho0=rho} holds, and making some
physically motivated approximations, we find that
the $\phi _{R1,fi}$
are determined up to a constant by the {\em local molecular field equations}: 
\begin{equation}
\phi _{R1,fi}({\bf r}_{i})=\int^{\prime }d{\bf r}_{j}[\delta (f,j)+\rho
_{R,fj}({\bf r}_{j})]\,u_{1,ji}({\bf r}_{ji}).  \label{lmforigin}
\end{equation}
Here the prime on the integral indicates an implicit summation over all
species $j$ and an integration over the angles of the body frames. Long
ranged interactions from the fixed species $f$ are accounted for by the
$\delta (f,j)$ term, which denotes products of $\delta $-functions describing
the fixed location and orientation of $f$.

Note that the average over the slowly varying $u_{1,ji}({\bf r}_{ji})$ in
Eq.~\ref{lmforigin} is weighted by $\rho _{R,fj}({\bf r}_{j})$, the {\em %
singlet} density for species $j$ (in the effective field of fixed species $f$
but with no explicit reference to species $i$). This neglect of pair
correlations between molecules at ${\bf r}_{j}$ and ${\bf r}_{i}$
characterizes a mean field approximation \cite{levin02}, and in most
contexts this would represent a major source of error. However, a general
feature illustrated by Fig.~\ref{potentialseparation} is
that as $\sigma $ increases the $u_{1,ji}$
become progressively more slowly varying at short distances. Thus we can
ensure that all the $u_{1,ji}$ will be slowly varying over the length scales
of relevant local pair correlations in the system by choosing $\sigma $
larger than some state dependent minimum value $\sigma _{\min }$. This
permits a consistent and controlled use of the mean field approximation in
computing the average, and we anticipate very accurate results from the LMF
theory for any $\sigma \geq \sigma _{\min }$ if we properly describe the
resulting density in the mimic system \cite{chen04}.

At strong coupling we argue that $\sigma _{\min }$ should be of order a
characteristic nearest neighbor spacing $\bar{a}.$ The strong short ranged
parts of the Coulomb interactions on the scale of $\bar{a}$ and below
directly affect pair correlations between nearest neighbor molecules. These
will be consistently described in the mimic system if we choose $\sigma
=\sigma _{\min }$ of order $\bar{a}$ so that there are essentially nearest
neighbor interactions between the effective Coulomb cores and the averaged
effects of the $u_{1,ji}$ from the further neighbors are slowly varying on
this scale. We illustrate below the utility of these ideas for simple models
of ionic solutions at strong coupling.\\
\\
{\bf Size-asymmetric primitive model}

A model of great current interest is the size-asymmetric primitive model
(SAPM) of ionic solutions, a fluid of oppositely charged hard spheres of
different sizes in a uniform dielectric continuum. The different hard sphere
diameters crudely account for the different core sizes of real cations and
anions, and there is an interesting and not well-understood dependence of
the critical temperature and critical density in this model as the size or
charge ratio is varied \cite{kim05}. We consider in particular the uniform equimolar
system studied using Monte Carlo (MC) simulations \cite{weis02} by Weis and Levesque
(WL), with symmetric unit charges $e_{0}=\bar{q}_{1}=-\bar{q}_{2}$ and a
diameter ratio $d_{1}/d_{2}=0.4$. Thus $w_{s,ij}(r)=\infty $ for $r\leq %
d_{ij}\equiv (d_{i}+d_{j})/2,$ and is zero otherwise, and $w_{q,ij}(r)=\bar{q%
}_{i}\bar{q}_{j}/\epsilon r$.

WL characterize the states by two dimensionless parameters: a reduced density
$\rho ^{*}\equiv (N_{1}+N_{2})d_{2}^{3}/V$ and an effective coupling strength
$\Gamma^{*}\equiv l_{B}/d_{2}$ (called $q^{*2}$ in
the notation of WL). Here $l_{B}\equiv \beta e_{0}^{2}/\epsilon $ is the {\em Bjerrum
length}, the distance where the interaction energy between two unit charges
$e_{0}$ equals $k_{B}T,$ and $\beta \equiv (k_{B}T)^{-1}$.
Their simulations indicate that   the critical point occurs at $\rho ^{*}=0.195$
and $\Gamma ^{*}=15.15.$ We report results here for two strong coupling
states with very different properties: a high density subcritical liquid
state with $\rho ^{*}=1.4$ and $\Gamma ^{*}=16$, and a low density
supercritical vapor state with $\rho ^{*}=0.04$ and $\Gamma ^{*}=9$.

The competition between the Coulomb interactions and the packing
arrangements of the embedded hard cores in the SAPM produces elaborate local
structures in these strong coupling states as exhibited in the pair
correlation functions $g_{ij}(r)$, proportional to the density response to
an external field $\phi _{ij}(r)=w_{ij}(r)$ arising from a fixed ion of type $i$ at the origin.
Thus in LMF theory even uniform fluid correlation functions are described from a
nonuniform point of view. These characteristic features can be very accurately
reproduced in the mimic system using the {\em strong coupling approximation}
(SCA). The SCA replaces $\phi _{R,ij}(r)$ by the known strong short
ranged component $\phi _{0,ij}(r)=u_{0,ij}(r)$ of the field from fixed ion
$i$. This corresponds to fixing a mimic particle at the origin, or
equivalently, approximating the $g_{ij}(r)$ in the uniform ionic system by
the $g_{0,ij}(r)$ in the uniform mimic system \cite{chen04}.

\begin{figure}[tbp]
\includegraphics[%
 width=1.0\columnwidth]{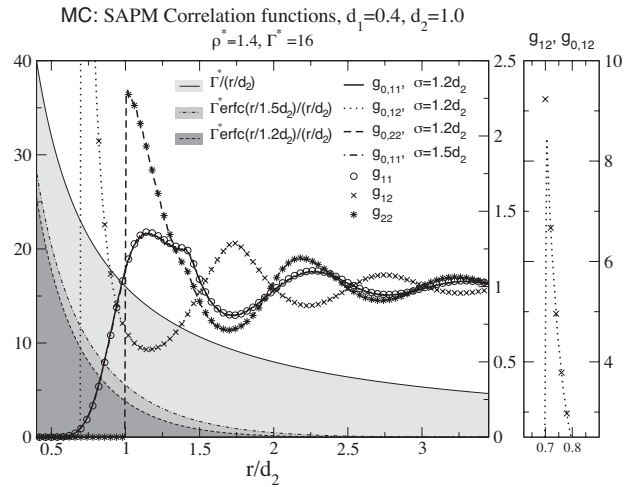} 
\caption{Dimensionless potentials and pair correlation functions
for the SAPM at high density and strong coupling.  The potentials use
the left vertical axis. Both the full potential between positive ions
(light grey shading) and the mimic interactions for two different
values of $\sigma $ (darker grey shadings) are shown.  The various pair
correlation functions use the right vertical axis. Results for
$g_{0,11}$  using two different values of $\sigma $ are shown; the
differences are barely visible on the scale of the graph. The right
inset focuses on the high first peak of the cation-anion correlation
function.}
\label{SAPMhigh}
\end{figure}

In Fig.~\ref{SAPMhigh} we compare correlation functions determined by WL for the high
density state with $\rho ^{*}=1.4$ and $\Gamma ^{*}=16$ to MC
simulations we carried out in the uniform mimic system with a ``molecular sized''
choice of $\sigma =1.2d_{2}$ using the SCA. Simulations of the long
ranged system required careful and costly treatment of periodic boundary
conditions using the Ewald sum method; this was not needed for the short
ranged mimic system. Despite the very different range and magnitude of the mimic
system interactions, all the pair correlation functions 
are strikingly similar to those of the full SAPM. These
functions are very different from the profiles of the associated hard sphere
mixture with the charges set equal to zero, indicating the crucial importance of including 
the strong short ranged parts of the Coulomb interactions $w_{q0,ij}$ in defining the
mimic interactions in Eq.~\ref{u0mimic}. Equally good results are found for
larger values of $\sigma $, as illustrated in the figure, but the good agreement
fails for much smaller
$\sigma $, indicating that $\sigma _{\min }$ is about $1.2d_{2}$ for this state.

\begin{figure}[tbp]
\includegraphics[%
  width=1.0\columnwidth]{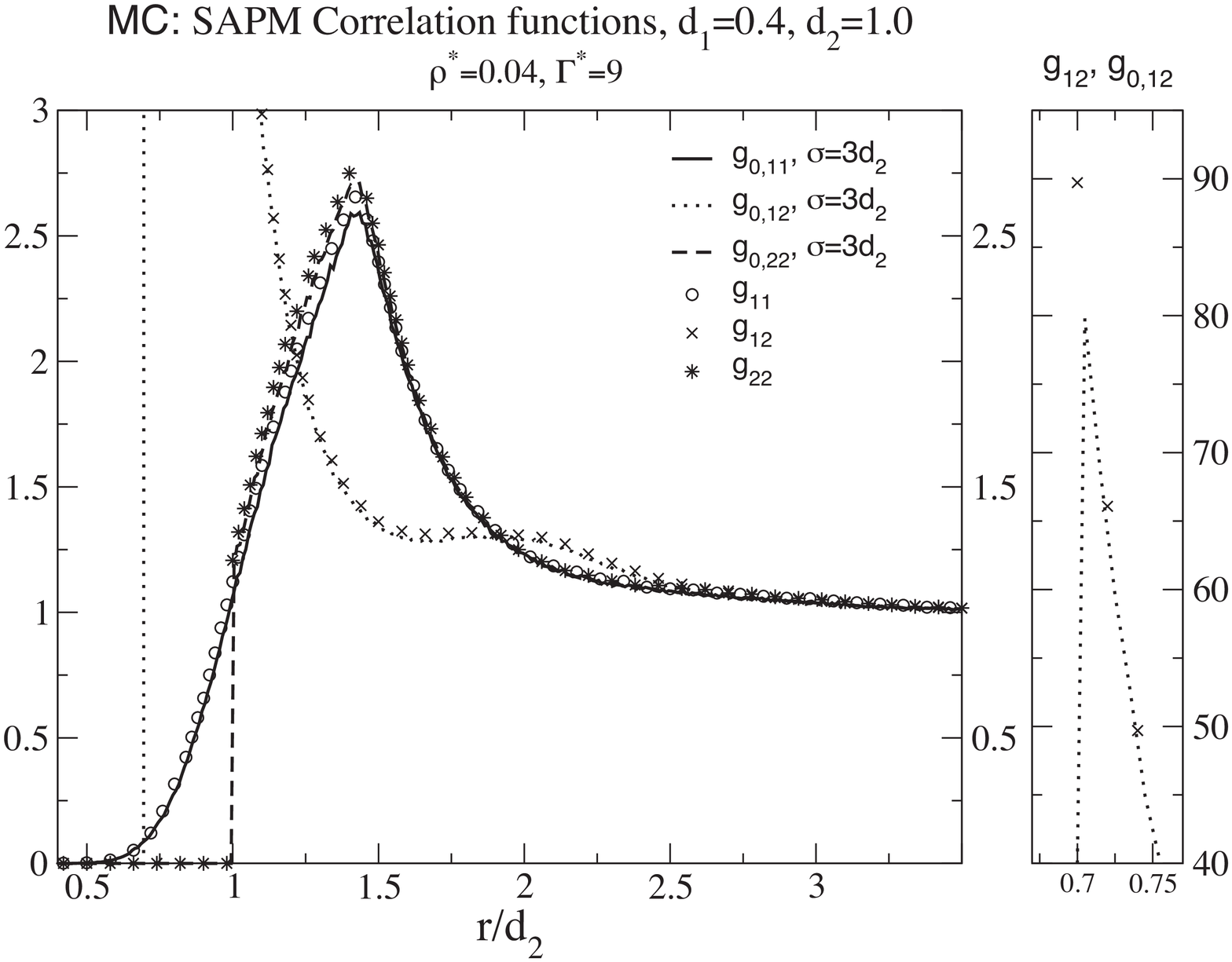}
  
\caption{Pair correlation functions for the SAPM in the low density ion pairing
regime, with the same conventions as in Fig.~\ref{SAPMhigh}.}
\label{SAPMlow}
\end{figure}

Qualitatively different structures are seen in the low density vapor state
with $\rho ^{*}=0.04$, and $\Gamma ^{*}=9$, as illustrated in Fig.~\ref{SAPMlow}. The
simulations of WL show that oppositely charged ions pair together with a
typical spacing close to the minimum permitted by the hard core diameters,
along with some transient formation of longer chain-like structures. The
correlation functions between like-charged pairs exhibit pronounced peaks of
essentially the same magnitude at a separation of $r=1.4d_{2}=d_{1}+d_{2}$.
This indicates the existence of small clusters of ion pairs, with the same
peak position and amplitude for both ``$+-+$'' or ``$-+-$''
configurations at the minimum distance permitted by a linear arrangement of
the embedded hard cores. These peaks also illustrate how counterions can induce an effective attraction between like charged objects, as discussed in
detail in the next section.
Very good agreement between full and mimic system
correlation functions is achieved with a choice of $\sigma _{\min}
=3.0d_{2}$, consistent with the larger average spacing between dilute ion pairs. 

The clustering of the ions has probably presented the most severe challenges
to theories of ionic systems. It is particularly crucial for the study of
critical phenomena and vapor-liquid coexistence \cite{kim05}. The PB
approximation and the frequently used hypernetted chain (HNC) integral
equation fail to predict ion clustering; indeed the HNC equation has no
solution in most of the ion pairing regime \cite{weis02}. In contrast, the
mimic system as described by the simple SCA already builds in the most
important local features of ion aggregation.  This very good agreement
strongly suggests that the LMF theory can accurately represent the 
Coulomb cores that contribute to local correlation functions in more realistic
models of ionic systems. Any remaining errors can be attributed mainly to
deficiencies in the description of the other short ranged core
interactions, thus permitting the efficient development of more accurate models.\\
\\
{\bf Charged walls with point counterions} 

The suspension and self-assembly of highly charged polyelectrolytes
(macroions) in the presence of mobile counterions (microions) is of great
interest in biological systems \cite{gelbart00}. These systems usually
involve charge and size asymmetries much greater than that of the SAPM and
are often studied by fixing a certain macroion configuration and computing
the microion distribution and resulting forces on the macroions. We discuss
here the simplest such model system \cite{Wennerstrom82}, consisting of
uniformly charged infinite hard walls with neutralizing point counterions
(and no co-ions) in a uniform dielectric environment. This model is simple
enough that exact results in certain limits are known \cite{Moreira00}, but
it still illustrates many fundamental issues that arise from the interplay
between long and short ranged forces in an explicitly nonuniform
environment. It is clear from the previous section that the LMF theory can
deal with more realistic models for the walls and counterions.\\
\\
{\bf One charged wall} We first consider the case of a single hard wall with
a uniform negative charge density $q_{w}$ at the $z=0$ plane, where we take
the zero of electric potential energy. Without loss of generality we can
assume that the counterions have a (unit) charge $e_{0}$ and express the
wall charge density $q_{w}\equiv -e_{0}/l_{w}^{2}$ in terms of the length
$l_{w}$ of the side of a square enclosing that amount of charge. The
potential energy $\phi ^{1w}(z)$ of a counterion at a distance $z$ from the
wall is $2\pi e_{0}^{2}z/(\epsilon l_{w}^{2})$. 
The {\em Gouy-Chapman length} $l_{G}$ is defined as the distance where 
this potential equals $k_{B}T$, {\it i.e.},
$l_{G}\equiv k_{B}T\epsilon l_{w}^{2}/(2\pi e_{0}^{2})=l_{w}^{2}/(2\pi l_{B})$.
$l_{G}$ characterizes the effective
strength of the attractive wall-counterion interaction, and most counterions
will be found near the wall in an effective slit whose width is proportional
to $l_{G}$. Dimensionless combinations of thermodynamic variables in this
simple system depend only on a single  control parameter \cite{Moreira02} $\xi
\equiv l_{B}/l_{G}=l_{w}^{2}/(2\pi l_{G}^{2})$.

As $\xi $ increases (e.g., by decreasing $T$ at fixed wall and counterion
charge) counterions are driven increasingly close to the wall by the
decreasing $l_{G}$. At strong coupling with $\xi \gg 1$ or $l_{G}\ll
l_{w} $, most counterions are next to the wall and form a
(``strongly-correlated'') two dimensional (2D) liquid layer \cite
{rouzina96,Shklovskii99} with average lateral spacing $\bar{a}\simeq %
a_{2D}\equiv l_{w}$ fixed by local neutrality. There are indeed strong lateral
correlations between the counterions in the 2D\ layer: the coupling strength 
$\Gamma _{a_{2D}}\equiv l_{B}/a_{2D}=\xi ^{1/2}/(2\pi )^{1/2}\gg 1$ for
large $\xi $. As discussed above we then expect the effective Coulomb core
size $\sigma _{\min }$ to be of order $a_{2D}=l_{w}$. However, because of
these repulsive cores, particles cannot stack perpendicular to the wall and
remain near the narrow slit. Thus the density outside the slit is very low
and there are only weak correlations normal to the wall.

In the opposite weak coupling limit with $\xi \ll 1$ or $l_{G}\gg l_{w}$
counterions can take advantage of the larger effective volume of the slit
and adopt a more diffuse 3D packing to reduce their repulsive interactions.
Crudely assuming all counterions are found within $l_{G}$ of the wall and
using a simple cubic lattice to estimate the characteristic counterion
spacing in this volume we now have $\bar{a}\simeq $ $a_{3D}\equiv
(l_{w}^{2}l_{G})^{1/3}=$ $l_{w}/(2\pi \xi )^{1/6}$. There is weak coupling
between the counterions, with
$\Gamma _{a_{3D}}\equiv l_{B}/a_{3D}=\xi ^{2/3}/(2\pi )^{1/3}\ll 1$
and here it is natural to take $\sigma _{\min }\simeq l_{B}=l_{w}(\xi
/2\pi )^{1/2}$ as an estimate for the effective Coulomb core size \cite{chen04}. The
crossover to strong coupling with essentially 2D packing and $\sigma _{\min
}\simeq $ $a_{2D}=l_{w}$ occurs for $\xi $ of order unity, and the 2D
packing indeed provides a larger average spacing at large $\xi $.

Quantitative results take an especially simple form \cite{Moreira02} if we
introduce a dimensionless rescaled density $n(z/l_{G})\equiv
l_{G}l_{w}^{2}\rho (z)$ that incorporates the anticipated ($\xi $%
-dependent)\ scaling of the profile with $l_{G}$. Local neutrality requires
that $\int_{0}^{\infty }d\tilde{z}n(\tilde{z})=1$, where $\tilde{z}\equiv
z/l_{G}$. Lengths scaled by $l_{G}$ will generally be indicated by a tilde.
Moreover, because of the impulsive $\delta $-function force at a hard wall,
there is an exact relation between the pressure and the contact density.
This yields the well-known {\em contact theorem}, which implies $n(0)=1$ for
the contact value of the rescaled density at a single charged hard wall \cite
{Wennerstrom82}.

Exact results \cite{Moreira00} for $n(\tilde{z})$ are known
in the limit $\xi \rightarrow 0$
from a rigorous weak coupling expansion, which gives results agreeing with
the PB approximation, $n_{PB}(\tilde{z})=1/(\tilde{z}+1)^{2}$. A different
strong coupling expansion gives exact results as $\xi \rightarrow \infty $,
$n_{SC}(\tilde{z})=e^{-\tilde{z}}$. However, attempts to connect these limits
by analyzing higher order terms in each expansion have had only limited success 
\cite{Moreira02}. We now show that the LMF theory provides a simple,
accurate, and unified approach for general $\xi $.\\
\\
{\bf LMF equation for one charged wall} We can take
advantage of planar symmetry and integrate exactly over the lateral degrees
of freedom in the long ranged parts $u_{1,ji}({\bf r}_{ji})$ of the
counterion-counterion and wall-counterion interactions in Eq.~\ref
{lmforigin}. The resulting LMF equation can be written in dimensionless
form for $\tilde{z}_{1},\tilde{z}_{2}\geq 0$ as 
\begin{equation}
\tilde{\phi}_{R1}(\tilde{z}_{1})=\int\nolimits_{0}^{\infty }d\tilde{z}%
_{2}[-\delta (\tilde{z}_{2})+n_{R}(\tilde{z}_{2})]G(\tilde{z}_{2},\tilde{z}%
_{1}).  \label{lmf1w}
\end{equation}
Here $\tilde{\phi}_{R1}(\tilde{z}_{1})\equiv \beta \phi _{R1}(\tilde{z}%
_{1}l_{G})$ and $G(\tilde{z}_{2},\tilde{z}_{1})\equiv -|\tilde{z}_{1}-\tilde{%
z}_{2}|$ erf$\left( |\tilde{z}_{1}-\tilde{z}_{2}|/\tilde{\sigma}\right) -\pi %
^{-1/2}\tilde{\sigma}e^{-\left[ (\tilde{z}_{1}-\tilde{z}_{2})/\tilde{\sigma}%
\right] ^{2}}+|\tilde{z}_{2}|$erf$\left( |\tilde{z}_{2}|/\tilde{\sigma}%
\right) +\pi ^{-1/2}\tilde{\sigma}e^{-[\tilde{z}_{2}/\tilde{\sigma}]^{2}}$
is the Green's function associated with a normalized planar Gaussian charge
distribution centered at $\tilde{z}_{2}$, with the zero of potential energy
at $\tilde{z}_{1}=0$.

The $-\delta (\tilde{z}_{2})$ term in Eq.~\ref{lmf1w} accounts for the
long ranged component $\tilde{\phi}_{1}(\tilde{z}_{1})=-G(0,\tilde{z}_{1})$
of the dimensionless attractive potential $\tilde{\phi}^{1w}(\tilde{z}_{1})=%
\tilde{z}_{1}$ between a counterion at $\tilde{z}_{1}$ and the negatively
charged wall at $\tilde{z}_{2}=0$. The remaining short ranged part $\tilde{%
\phi}_{0}(\tilde{z}_{1})=\tilde{z}_{1}-\tilde{\phi}_{1}(\tilde{z}_{1})$ of
the wall potential is 
\begin{equation}
\tilde{\phi}_{0}(\tilde{z}_{1})=\tilde{z}_{1}\text{erfc}(\tilde{z}_{1}/%
\tilde{\sigma})-\tilde{\sigma}e^{-(\tilde{z}_{1}/\tilde{\sigma})^{2}}/\sqrt{%
\pi }+\tilde{\sigma}/\sqrt{\pi }.  \label{phi01w}
\end{equation}
The effective field is then $\tilde{\phi}_{R}(\tilde{z}_{1})=\tilde{\phi}%
_{0}(\tilde{z}_{1})+\tilde{\phi}_{R1}(\tilde{z}_{1}),$ with $\tilde{\phi}%
_{R1}$ given by Eq.~\ref{lmf1w} for $\tilde{z}_{1}\geq 0,$ and infinity
otherwise.\\
\\
{\bf mPB approximation} To solve Eq.~\ref{lmf1w} self consistently we must
accurately determine the density $n_{R}(\tilde{z})$ induced by $\tilde{\phi}%
_{R}(\tilde{z})$. At weak coupling, neighboring ions interact weakly and the
density response to the effective field is proportional to the ideal gas
Boltzmann factor $\exp [{-\tilde{\phi}_{R}(\tilde{z})}]$. Using this
approximation in Eq.~\ref{lmf1w}, we have a closed equation, which we
refer to as the {\em mimic Poisson-Boltzmann} (mPB)\ equation. Moreover, we
can show \cite{chen04a} for all $\xi $ that a self consistent solution of
the mPB equation will exactly satisfy both neutrality and the contact
theorem. The latter implies that the density response takes the simple form
$n_{R}(\tilde{z})=\exp [{-\tilde{\phi}_{R}(\tilde{z})}]$ with our choice of zero
of energy.

Remarkably, however, the mPB approximation also gives accurate results at 
{\em strong coupling} with $\xi \gg 1$, where there is an essentially 2D
arrangement of the mimic particles in an effective narrow slit. Correlations
normal to the wall are very weak and the Boltzmann factor again can
accurately describe the density response to the $z$-dependent field $\tilde{%
\phi}_{R}(\tilde{z})$, as can be verified by more formal arguments \cite
{Moreira02,chen05}.

\begin{figure}[tbp]
\includegraphics[%
 width=1.0\columnwidth]{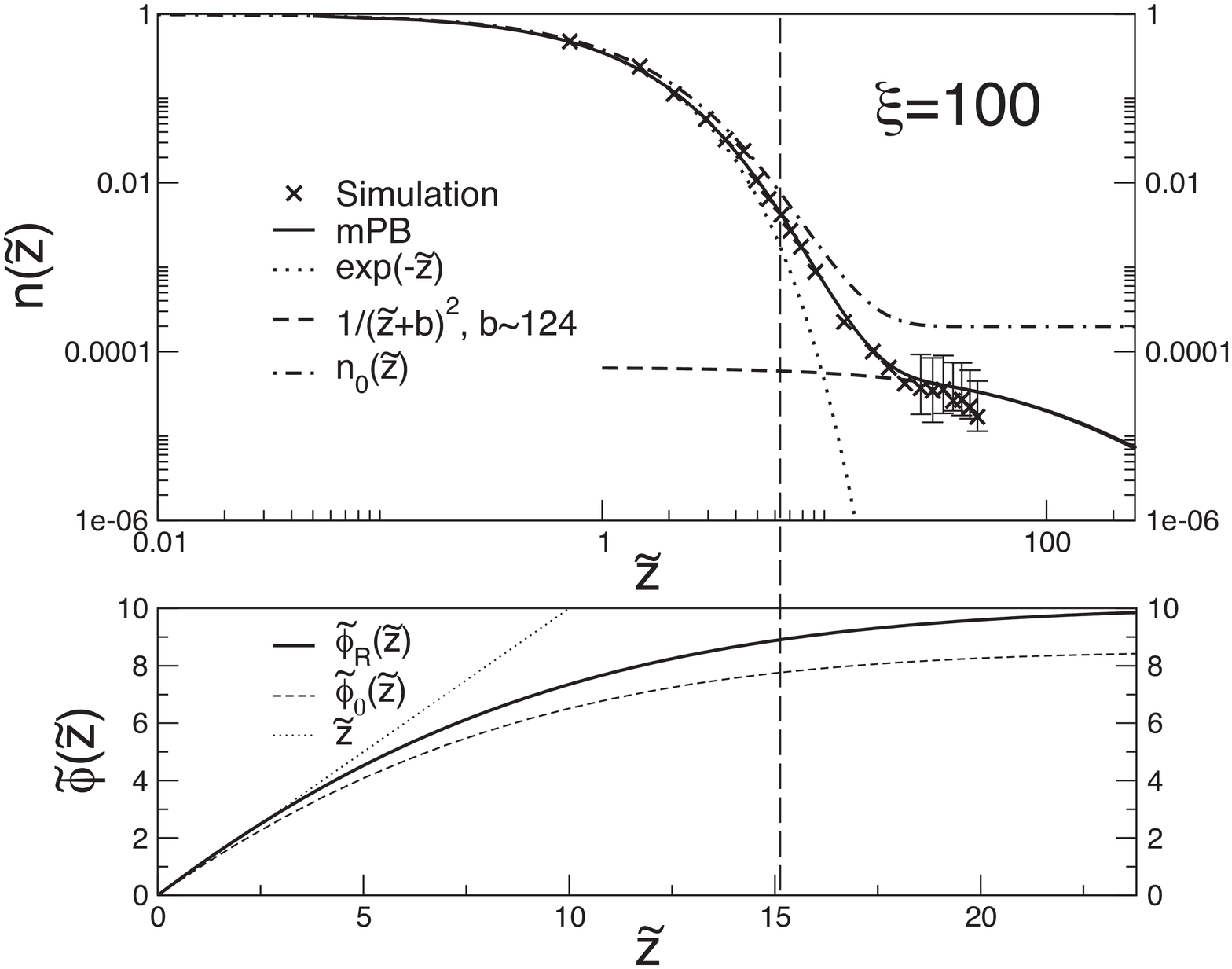}
\caption{Upper graph: (Note the log scales) Rescaled counterion density near one
planar charged wall calculated 
by the full mPB theory and by the SCA compared to computer simulation
data in \cite{Moreira02}. The limiting exponential profile provides a good fit
only near the wall. We can show analytically \cite{chen04a} that there is a
crossover to an algebraic tail of the form $1/(\tilde{z}+b)^2$ in the mPB theory
for $\tilde{z}\simeq \tilde{\sigma }^{1w}$ as seen in the graph. The vertical
dashed line in both graphs indicates the value of $\tilde{\sigma }^{1w}$.
Lower graph: (Note the linear scales) The full dimensionless wall potential $\tilde{z}$
compared to $\tilde{\phi}_{R}(\tilde{z})$ and
$\tilde{\phi}_{0}(\tilde{z})$ from the mPB theory.}
\label{1wall}
\end{figure}

This motivates our use of the mPB approximation with $n_{R}(\tilde{z})=
\exp [{-\tilde{\phi}_{R}(\tilde{z})}]$ for all $\xi $ in
Eq. \ref{lmf1w}. The mPB approximation is least justified at intermediate
values of $\xi $, and it breaks down if $\sigma $ is chosen much larger than 
$\sigma _{\min }$ so that there would be strong direct interactions
between further neighbors in the mimic system. Thus we provide a smooth
interpolation between the known limiting values of $\sigma _{\min }$ in the
weak and strong coupling regimes by choosing $\sigma =\sigma ^{1w}\equiv
C\min (l_{B},l_{w})$, and fix $C=0.60$ by finding the best fit to
simulations \cite{Moreira02} at a moderately strong coupling state with
$\xi=40$. In this example it is numerically more convenient to differentiate
the resulting mPB equation and solve for the effective force, which vanishes
far from the wall, and then
get the effective field by integration \cite{rodgers05}. An iterative solution is
straightforward and no other simulation data are required. \\
\\
{\bf Results for one charged wall} Figure \ref{1wall} gives results for $n_{R}(%
\tilde{z})$ at strong coupling with $\xi =100$. There is excellent agreement
between the results of the mPB theory and MC simulations \cite{Moreira02} of
the long ranged system carried out by Moreira and Netz. The log-log
plot emphasizes that $n_{R}(\tilde{z})$ has two characteristic regions. Near
the wall there is an initial exponential decay arising mainly from particles
in the 2D layer. This continues until about $\tilde{z}\simeq \tilde{\sigma}%
_{\min }/2$ where the density is very low and there is a crossover to
algebraic decay as in the PB solution, but with a much larger effective
$l_{G}$. A natural physical interpretation is that the
small fraction of counterions outside the 2D layer adopt a diffuse
PB-like profile generated by an effective wall whose charge density
has been greatly reduced by the charge of the
tightly bound counterions. This idea has been suggested
before \cite{Shklovskii99}, but LMF theory is the first unified theory capable of
describing both limiting regions and the crossover region.

The density near the wall is very accurately described by the even simpler
SCA, where $\tilde{\phi}_{R}(\tilde{z})$ is approximated by $\tilde{\phi}%
_{0}(\tilde{z})$. The resulting density $n_{0}(\tilde{z})\equiv \exp [{-\tilde{\phi}_{0}(%
\tilde{z})}]$ can be written down immediately from Eq.~\ref{phi01w}.
As shown in Fig.~\ref{1wall} both $\tilde{\phi}_{0}(\tilde{z})$
and $\tilde{\phi}_{R}(\tilde{z})$ closely resemble the full potential
$\tilde{z}_{1}$ near the wall for $\tilde{z}_{1}\lesssim \tilde{\sigma}_{\min}/2$.
But $n_{0}(\tilde{z})$ cannot describe the PB-like region at large
$\tilde{z}$ as does the full mPB theory and it does not obey the neutrality
condition. This example highlights both the strengths and weaknesses of the
SCA. When properly used to describe only short ranged correlations at strong
coupling very accurate results can be found.\\
\\
{\bf Two charged walls} We now briefly consider two equally charged
hard walls forming a real slit with width $d$, with neutralizing point
counterions in between. At strong coupling and intermediate widths, the
counterions can induce an effective attractive force between the walls. Such
effective attraction between like charged objects may have important
experimental consequences, and it has generated a great
deal of theoretical interest \cite{gelbart00, levin02,grosberg02,boroudjerdi05}.

Reference \cite{Wennerstrom82} gives an exact expression for the
dimensionless (osmotic) pressure $\tilde{P}\equiv \beta l_{G}l_{w}^{2}P$
arising from neutralizing point counterions confined between charged hard
walls at $z=0$ and $z=d$: 
\begin{equation}
\tilde{P}=n(0)-1.  \label{P osmotic}
\end{equation}
Thus if the rescaled contact density $n(0)$ is less than (greater than) one
there is an effective attractive (repulsive) force on the walls. As $\
d\rightarrow \infty $, we recover the one-wall results discussed earlier,
where $\tilde{P}=0$, and $n(0)=1$.

Since the total force on a counterion from equally charged walls exactly
cancels for all $z$ and all $d$, we now have $\tilde{\phi}^{2w}(\tilde{z})=0$
for $0\leq \tilde{z}$ $\leq \tilde{d}$. As in the one-wall case, it is
useful to divide $\tilde{\phi}^{2w}$ into a short ranged part 
\begin{equation}
\tilde{\phi}_{0}^{2w}(\tilde{z};\tilde{d})\equiv \tilde{\phi}_{0}^{1w}(%
\tilde{z})+\tilde{\phi}_{0}^{1w}(\tilde{d}-\tilde{z})-\tilde{\phi}_{0}^{1w}(%
\tilde{d}),  \label{phi02w}
\end{equation}
given by a sum of short ranged single-wall terms (indicated by the
superscript $1w$) defined in Eq.~\ref{phi01w}, and the remainder. We take
the zero of energy on the left wall at $\tilde{z}=0.$ The effective field
$\tilde{\phi}_{R}(\tilde{z})=$ $\tilde{\phi}_{0}^{2w}(\tilde{z};\tilde{d})+%
\tilde{\phi}_{R1}(\tilde{z})$ is determined from the two-wall LMF equation.
This closely resembles Eq.~\ref{lmf1w}, except that the integration is
from $0$ to $\tilde{d}$ and there is an additional $-\delta (\tilde{d}-%
\tilde{z}_{2})$ term in the integrand, accounting for interactions with the
second wall at $z=d$. Again we use the mPB approximation $n_{R}(\tilde{z}%
)=A\exp [{-\tilde{\phi}_{R}(\tilde{z})}]$ and fix the constant $A$ (which equals
the contact density with our choice of zero of energy) using the neutrality
condition $\int_{0}^{\tilde{d}}d\tilde{z}n_{R}(\tilde{z})=2$. The pressure
is then given by Eq.~\ref{P osmotic}.

The resulting two-wall mPB equation reduces exactly to an integrated
form of the PB equation if $\sigma =0$. The latter has an analytic
solution and predicts a repulsive force for all $d$ and $\xi $ \cite{levin02,Moreira02}.
The mPB theory also predicts a weak repulsive force at strong coupling for
sufficiently large $d$, arising from weak repulsions between counterions
in the dilute PB-like tails of the one-wall profiles discussed above. On the
other hand, at strong coupling and sufficiently small $d$, core repulsions
make it unfavorable for particles in the narrow slit to stack
perpendicular to the walls, and
counterions will be forced into a {\em single} 2D layer with characteristic
lateral spacing $\bar{a}\backsimeq l_{w}/\sqrt{2}$ fixed by neutrality. To
interpolate between this limit and weak coupling we choose $\sigma =\sigma
^{2w}\equiv C\min (l_{B},l_{w}/\sqrt{2})$, and take the same value of
$C=0.60 $ that we used for the one-wall theory.
Since $\tilde{\sigma}_{\min }=C\xi $ at small $\xi $, the
PB approximation is consistent only as $\xi \rightarrow 0$.
The mPB theory naturally introduces a crucial new length scale $\sigma _{\min }$
that allows for a change in the functional form of $n_{R}(\tilde{z})$ as $\xi $
increases.

\begin{figure}[tbp]
\includegraphics[%
width=1.0\columnwidth]{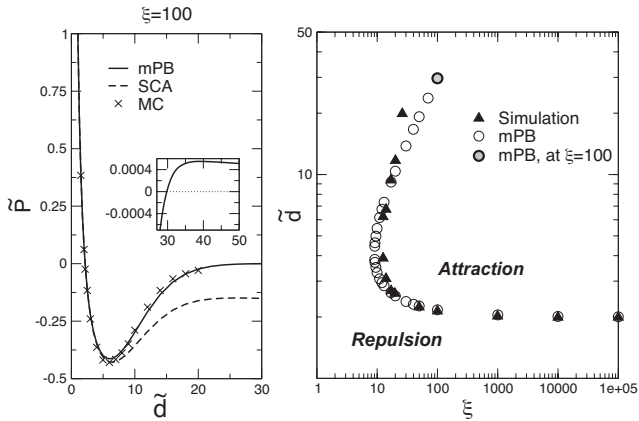}  
\caption{Left graph: Dimensionless pressure at strong coupling
between two equally charged hard walls as a function of width $\tilde{d}$
from \cite{Moreira02} compared to predictions from the full mPB theory and
the SCA. The inset shows that the mPB theory predicts a weak repulsive
force at larger widths. Right graph: Repulsive and attractive forces between
two charged walls as as a function of width and coupling strength as
determined by MC simulations and the mPB theory.}
\label{2walls}
\end{figure}

Figure \ref{2walls} compares numerical results of mPB theory \cite{rodgers05} to
simulation data \cite{Moreira02} for strong coupling states. The left
graph shows for $\xi =100$ there is very good agreement between the mPB
theory and computer simulations for all widths where
simulations can be performed. As shown in the inset, the mPB theory predicts
that at still larger widths there is a weak repulsive force between the
walls. This reentrant behavior is illustrated more generally in the right
graph, which also shows that a minimal coupling strength of $\xi \geq \xi
_{c}\simeq 12$ is needed to get any attraction \cite{Moreira02}.

The left graph also shows
results from the analytic SCA, where $n(\tilde{z})$ is approximated by
$n_{0}(\tilde{z})\equiv A_{0}\exp [{-\tilde{\phi}_{0}^{2w}(\tilde{z};\tilde{d})}]$,
with $A_{0}$ similarly determined by neutrality. At strong coupling and
large $d$, $\tilde{\phi}_{0}^{2w}(\tilde{z};\tilde{d})$ has a deep
attractive well near each wall essentially identical to that found near a
single wall. However at small enough $d$ the wells from the individual
$\tilde{\phi}_{0}^{1w}\ $terms in Eq.~\ref{phi02w} begin to overlap and
their depth decreases. As $\ d\rightarrow 0$, $\tilde{\phi}_{0}^{2w}(\tilde{z%
};\tilde{d})$ vanishes for all $\tilde{z}$ in the slit. The SCA fails at
large widths, just as it did far from the wall in the one-wall case, and the
reentrant behavior is completely missed. However, it is very accurate at
smaller widths and it gives a good description of the location and magnitude
of the maximum attractive force.

The formation of the single 2D layer at sufficiently small widths $\tilde{d}<%
\tilde{\sigma}^{_{2w}}$ plays a key role in producing a strong
attractive force. Because of the absence of correlations normal to the walls, the
density profile will be relatively constant across the slit. We can use the
SCA to describe several features analytically in this regime. The minimum
pressure $\tilde{P}^{*}$ should occur near the largest width
$\tilde{d}^{*}\equiv 2\tilde{z}_{m}^{*}$ for which the single 2D layer remains
stable, defined by $\tilde{\phi}_{0}^{2w}(\tilde{z}_{m}^{*};\tilde{d}^{*})=1$.
At larger widths there will be higher contact densities as separate 2D
layers at each wall begin to form, and the nearly constant profiles at 
smaller widths will have higher contact densities by
normalization. Expanding $\tilde{\phi}_{0}^{2w}(\tilde{z};\tilde{d})$ in a
Taylor series, we have $\tilde{\phi}_{0}^{2w}(\tilde{z};\tilde{d})=2\tilde{z}%
(\tilde{d}-\tilde{z})/(\pi ^{1/2}\tilde{\sigma}^{_{2w}})$ to lowest order;
higher order terms are negligible for $\tilde{d}\ll \tilde{\sigma}^{_{2w}}$.
This implies $\tilde{d}^{*}=(2\pi ^{1/2}\tilde{\sigma}^{_{2w}})^{1/2}\simeq
1.94\xi ^{1/4}$, using our expression for $\tilde{\sigma}^{_{2w}}$.
Similarly evaluating $A_{0}$, the minimum pressure is
$\tilde{P}^{*}\simeq 3.717/\tilde{d}^{*}-1$. When $\tilde{P}^{*}= 0$
there can be no attractive forces; this determines  the minimal
coupling strength $\xi _{c}\simeq 13.48$ and associated critical spacing
$\tilde{d}_{c}^{*}\simeq 3.717$ needed to get any attraction. Finally, for
$\xi \gg \xi _{c}$ and $\tilde{d}=2$ we see $\tilde{\phi}_{0}^{2w}(\tilde{z};%
\tilde{d})\ll 1$, so the constant profile $n_{0}(\tilde{z})=2/\tilde{d}$ is
very accurate. Equation \ref{P osmotic} then implies that at very strong
coupling the transition from strong repulsive to strong attractive forces occurs
near $\tilde{d}=2$, as shown in \cite{Moreira00,Moreira02}. All these
predictions are in very good agreement with numerical solutions of the mPB
theory (and with simulations of the full and mimic systems, where
available) for all strong coupling states, as illustrated in Fig.~\ref{2walls}.\\
\\
{\bf Discussion}

In strong coupling regimes the short ranged parts of the Coulomb
interactions efficiently compete with the other short ranged molecular core
interactions and strongly influence pair correlations between neighboring
molecules. In LMF theory the choice of $\sigma _{\min}$ determines the
strength and range of these important Coulomb core interactions.
They play a key role in inducing an effective attraction between like-charged
objects at strong coupling, as illustrated here in Fig.~\ref{SAPMlow}
for the correlation functions between like-charged ions in the SAPM, and
in Fig.~\ref{2walls} for the osmotic pressure on
two like-charged walls.  In both cases, strong short ranged forces mainly
involving single counterions or a single counterion layer can mediate a
strong effective attraction. Such phenomena
have traditionally been interpreted as illustrating the ``breakdown of
mean field theory" and the need for new and more sophisticated
approaches. But LMF theory using the simple strong coupling
approximation, where only the short ranged parts of the Coulomb
interactions are taken into account, provides very accurate results at small and
moderate separations. At weak coupling and for long wavelength correlations
the averaged effects of the long ranged interactions as determined by
the LMF equation are needed as well \cite{chen04}.

LMF theory provides a general
conceptual framework that connects and clarifies
previous work on systems with both short and
long ranged interactions.  It has been used to describe
liquid-vapor interfaces and wetting and drying transitions for simple
fluids \cite{weeks02} and hydrophobic interactions in
water \cite{lum99}. It suggests new
and simplified simulation models for general Coulomb systems
based on the mimic system that do not require special
treatment of periodic boundary conditions. The short
ranged intermolecular interactions in the Coulomb mimic system are
reminiscent of the truncated interactions used in reaction field methods
\cite{hummer94}. But it was not very
clear in those approaches how to choose an appropriate cutoff and how
to treat nonuniform environments. The determination of $\sigma_{\min}$
and the effective field in LMF theory provides a way to deal with both problems.
We believe the LMF picture will prove useful not
only in formal theory, but also for qualitative reasoning and in
detailed simulations of biophysical systems.

Detailed results for charged hard walls, including MC simulations of the nonuniform
mimic system using both the SCA and the full LMF theory, with important contributions
made by Charanbir Kaur and Jocelyn Rodgers,
will be presented elsewhere \cite{rodgers05}.
We are grateful to them and to Michael Fisher for many helpful remarks.
We thank J. J. Weis for sending us his simulation data for the SAPM correlation
functions. This work was supported by the National Science Foundation
through grants CHE01-11104 and CHE05-17818.

\end{document}